\documentclass[preprint]{aastex}
\usepackage{natbib}
\usepackage{mathrsfs}
\usepackage{rotating}
\usepackage{graphicx}
\usepackage{epstopdf}
\usepackage{longtable,lscape}

\newcommand{\ud}{\mathrm{d}}

\newcommand{\fn}[1]{\footnote{\scriptsize{#1}}}
\newcommand{\Eqn}[1]{Eq{#1}.}  
\newcommand{\Fig}[1]{Fig{#1}.}  
\newcommand{\Cassit}{\textit{Cassini}}  

\slugcomment{Accepted to \textit{Icarus}}

\shorttitle{}
\shortauthors{Tiscareno, Thomas, and Burns}

\begin{document}

\title{\vspace{-0.15in}The Rotation of Janus and Epimetheus}
\author{Matthew~S.~Tiscareno,$^1$ Peter~C.~Thomas,$^2$ and Joseph~A.~Burns$^{1,3}$}
\affil{$^1$Department of Astronomy, $^2$Center for Radiophysics and Space Research, and \\$^3$Department of Theoretical and Applied Mechanics, Cornell University, Ithaca, NY 14853}

\begin{abstract}
Epimetheus, a small moon of Saturn, has a rotational libration (an oscillation about synchronous rotation) of $5.9^\circ \pm 1.2^\circ$, placing Epimetheus in the company of Earth's Moon and Mars' Phobos as the only natural satellites for which forced rotational libration has been detected.  The forced libration is caused by the satellite's slightly eccentric orbit and non-spherical shape. 

Detection of a moon's forced libration allows us to probe its interior by comparing the measured amplitude to that predicted by a shape model assuming constant density.  A discrepancy between the two would indicate internal density asymmetries.  For Epimetheus, the uncertainties in the shape model are large enough to account for the measured libration amplitude.  For Janus, on the other hand, although we cannot rule out synchronous rotation, a permanent offset of several degrees between Janus' minimum moment of inertia (long axis) and the equilibrium sub-Saturn point may indicate that Janus does have modest internal density asymmetries. 

The rotation states of Janus and Epimetheus experience a perturbation every four years, as the two moons ``swap'' orbits.  The sudden change in the orbital periods produces a free libration about synchronous rotation that is subsequently damped by internal friction.  We calculate that this free libration is small in amplitude ($<0.1^\circ$) and decays quickly (a few weeks, at most), and is thus below the current limits for detection using \Cassit{} images.
\end{abstract}

\textit{Subject headings:}  satellites, dynamics --- satellites, shapes --- Saturn, satellites \\
\indent{}\textit{Running header:}  Rotation of Janus and Epimetheus

\section{Introduction \label{Intro}}

\begin{figure}[!t]
\begin{center}
\includegraphics[width=10cm]{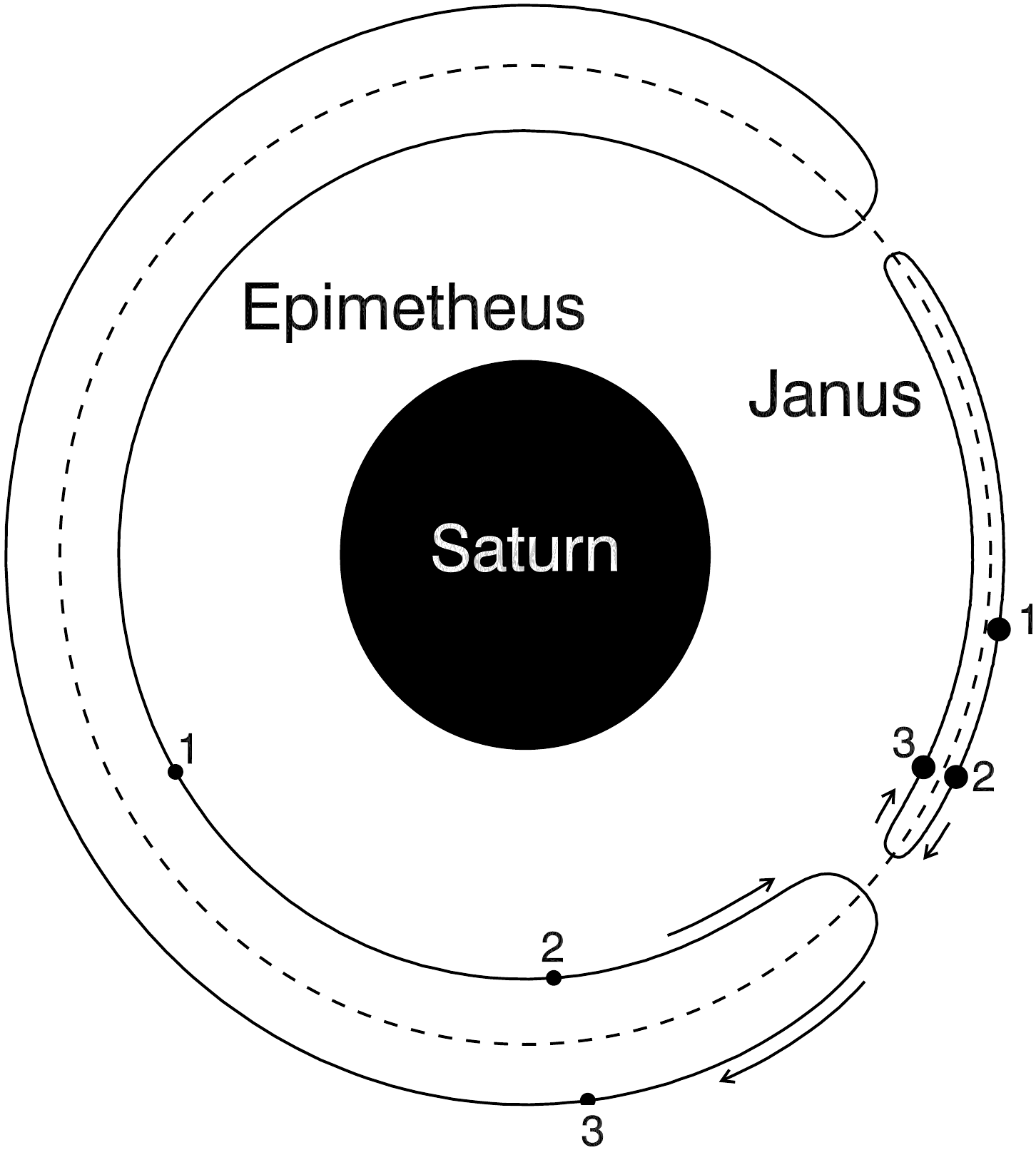}
\caption{The orbits of Janus and Epimetheus (with epicycles removed) in a frame rotating with their mass-averaged mean motion.  The moons' radial excursions are exaggerated by a factor of 500, and the moons' sizes are exaggerated by a factor of 50, but otherwise the figure is to scale.  Numbered points mark the moons' positions on 1)~2004~July~1, 2)~2005~May~21, and 3)~2006~September~9.  Arrow lengths indicate motion accomplished in 100 days.  \label{jeorbits8}}
\end{center}
\end{figure}

Janus and Epimetheus, two small satellites of Saturn, share a remarkable orbital configuration.  They move along the same orbit, but in contrast with other, much larger, bodies that are known to have small ``Trojans'' orbiting at their Lagrange points (including Jupiter, Mars, Neptune, Tethys, and Dione), neither Janus nor Epimetheus is overwhelmingly large compared to the other --- Janus is 3.6 times as massive as Epimetheus \citep{Jake08}.  Consequently, both Janus and Epimetheus (to which we refer collectively as the ``co-orbital moons'') execute ``horseshoe'' trajectories about their mass-averaged mean orbital state \citep{DM81a,DM81b,Yoder83,Peale86,jemodelshort}.  Every 4.00~yr they reach their mutual closest approach and ``swap'' orbits (\Fig{}~\ref{jeorbits8}) --- one moon's orbital rate (mean motion) slows down, which is to say that its mean distance from the planet (semimajor axis) increases, while the other moon does the opposite.  The effect on Epimetheus is greater than the effect on Janus, in proportion to their masses.  The most recent orbit swap occurred on 2006~January~21, at which time Janus became the inner satellite and Epimetheus the outer.  The next orbit swap will occur on 2010~January~21. 

The periodic change in the orbital rate provides a unique opportunity to observe the rotational response of the moons.  \citet{jemodelshort} made similar use of nature's obliging variation of a single parameter in a complex problem, observing morphological changes in the spiral density waves raised in Saturn's rings by Janus and Epimetheus and connecting them with the periodic changes in the orbital rate. 

If one assumes that the co-orbital moon is rotating synchronously (keeping the same face always towards Saturn) as it goes into the orbit swap, the change in the orbital rate means that the rotation rate is no longer synchronized.  Because Janus and Epimetheus are both significantly non-spherical, non-synchronous rotation does not continue indefinitely but results in simple harmonic motion (libration) about the most stable state in which the moon's long axis is pointed towards Saturn.  This kind of motion, which results from a one-time impulse (the orbit swap) and then decays with time due to friction in the moon's interior, is a \textit{free libration}.  In principle, observing the free libration's rate of decay can yield insight into the moon's interior properties. 

There are, however, two other forms of libration that must be considered, both of which arise due to eccentricity of the moon's orbit.  Firstly, because an eccentric moon moves faster at periapse and slower at apoapse, even a synchronously-rotating moon oscillates relative to the line connecting the moon and the planet (the \textit{optical libration}).  Secondly, the optical libration allows the planet to exert a torque on a non-spherical moon, resulting in a \textit{forced libration}.  The optical and forced librations are both in phase with the moon's orbit, with the moon's long axis pointing towards the planet at periapse and apoapse.  To date, the only known moons for which a forced libration has been measured are Earth's Moon \citep{Koziel67,Williams73} and Mars' Phobos \citep{Burns72,DC81,DC89,Duxbury89,Duxbury91,Simonelli93,Willner08}.  With this paper, we add Epimetheus to that number. 

In the next section, we will begin by considering tidal de-spinning.  We will then discuss the rotational free libration in the undamped (Section~\ref{Undamped}) and damped (Section~\ref{Damped}) cases, and then the rotational forced libration (Section~\ref{Forced}).  In Section~\ref{Data} we will compare our models with high-resolution imaging by the \Cassit{} Imaging Science Subsystem (ISS).  Section~\ref{Conclusions} presents a summary and conclusions. 

\section{The Na\"ive Case \label{Naive}}

\begin{figure}[!t]
\begin{center}
\includegraphics[width=10cm]{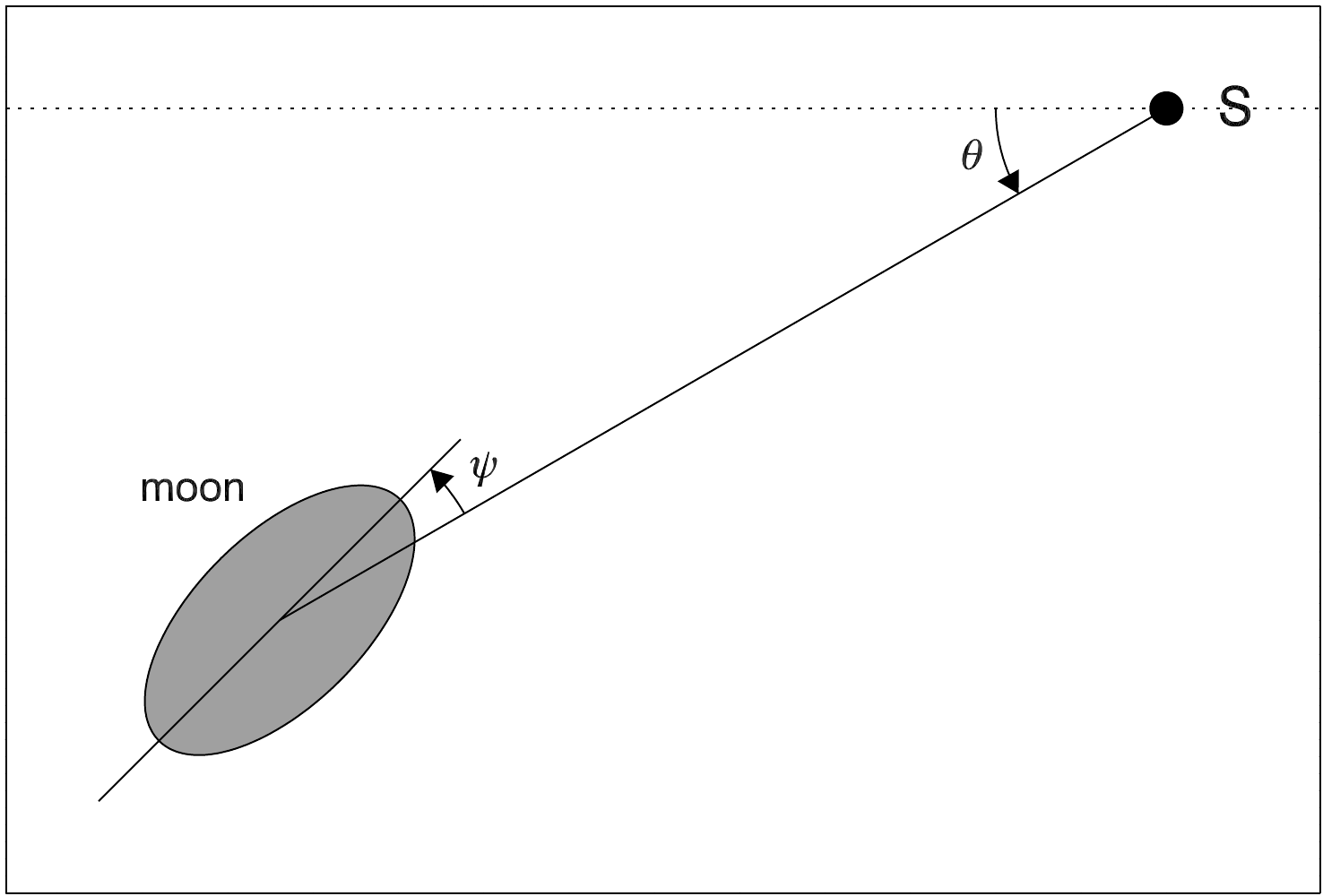}
\caption{The orientation $\psi$ of the moon's long axis is measured from the direction towards Saturn, while $\theta$ is the orbital longitude.  All angles are positive in the counter-clockwise direction.  Only in Section~\ref{Forced} is this picture complicated by non-circular orbits.  \label{coords_fig1}}
\end{center}
\end{figure}

In its simplest form, the problem is characterized by the orbital longitude $\theta$ and the libration angle $\psi$ (\Fig{}~\ref{coords_fig1}).  The orientation of the moon in inertial space is $\theta+\psi$.  In the models presented here, we will make the simplifying assumption that the orbit swap takes place instantaneously, with the mean motion changing from $n_1$ to $n_2$ at time $t=0$.  Since the moon was assumed to be rotating in a steady-state before the swap, $\dot{\psi}(t<0) = 0$, the pre-existing rotation rate of the moon in inertial space is $n_1$.  The orbit swap then results in an impulse $\dot{\psi}_i = n_1 - n_2$.  In fact, the orbit swap is not instantaneous, but takes place over a period of several months, making our estimate of $\dot{\psi}_i$ an upper limit.  

The standard treatment of tidal de-spinning of a fast rotator is to consider a torque exerted by the central planet on a tidal bulge raised by the planet but offset from the moon-planet line due to tidal dissipation, resulting in the expression \citep{Peale99}
\begin{equation}
\ddot{\psi} = - \frac{45}{76} \frac{\rho n^4 R^2}{\mu Q} ,
\end{equation}
where $\ddot{\psi}$ is negative iff $\dot{\psi} > 0$.  The density $\rho$, mean orbital rate $n$, and mean radius $\bar{R}$ for Janus and Epimetheus are given in Table~\ref{JEInfo3}. 
\begin{table}[!b]
\begin{scriptsize}
\caption{Parameters for tidal de-spinning \label{JEInfo3}}
\begin{tabular} { l c c c c }
\hline
\hline
 & $\bar{R}$ (km) & $m$ (10$^{18}$ kg) & $\rho$ (g cm$^{-3}$) & $\bar{n}$ ($^\circ$/dy) \\
\hline
Janus & 89.5 & 1.897 & 0.632 & 518.292 \\
Epimetheus & 58.1 & 0.526 & 0.641 & 518.292 \\
\hline
\end{tabular}
\\Note: $\bar{R}$ taken from Table~\ref{pct_params} below, while $m$ and $\bar{n}$ are taken from \citet{Jake08}; $\rho = 3m / 4 \pi \bar{R}^3$.
\end{scriptsize}
\end{table}

The ``quality factor'' $Q$ corresponds to the number of cycles it takes for rotational energy to substantially dissipate.  A common assumption for a monolithic block of ice is $Q \sim 100$, though $Q$ is likely much less for a fractured ``rubble-pile'' in which damping is made more efficient by increased internal friction.  We will use $Q \sim 10$ to represent the latter case.  The rigidity $\mu$ is commonly taken to be $\sim 4 \times 10^9$~N~m$^{-2}$ for solid ice, but also decreases by an order of magnitude or more for a fractured body \citep{GS09}.  

The resulting values of $\ddot{\psi}$ are small --- on the order of 1$^{\prime\prime}$~dy$^{-2}$ 
for solid ice, and some 100~times greater for a fractured rubble-pile.  The de-spinning time for a fast rotator (say, $\dot{\psi}_i = 100^\circ$~dy$^{-1}$) would thus be $\dot{\psi}_i/\ddot{\psi} \sim 1000$~yr (solid) or 10~yr (fractured).  While these de-spinning times are appropriate for the original damping of any primordial rotation of these moons, they are over-estimated for the case of the relatively small non-synchronous rotation induced by the orbit swap, $\dot{\psi}_i \sim 0.1^\circ$~dy$^{-1}$ (see Table~\ref{JEInfo}, and recall from the beginning of this Section that this is an upper limit), which leads instead to $\dot{\psi}_i/\ddot{\psi} \sim 1$~yr (solid) or 4~dy (fractured).  

We would then appear to be in a felicitous regime for which solid and fractured bodies might easily be distinguished, as non-synchronous rotation induced by an orbit swap is quickly damped for the latter but may remain for the entire 4-yr horseshoe libration for the former.  \textit{However,} the above analysis suffers from two faulty assumptions:  the first is that tidal de-spinning is controlled by the torque on the Saturn-raised tidal bulge, and the second is that de-spinning proceeds by slowing down a monotonic non-synchronous rotation.  Because Janus and Epimetheus in fact are not spherical but significantly triaxial (Table~\ref{JEInfo}), the rotational dynamics are controlled rather by the torque on the moon's figure.  Furthermore, because the moon's intrinsic figure does not change as the moon rotates, and because $\dot{\psi}_i$ is not large, the moon executes a \textit{libration} about the synchronous rotation state, with tidal effects appearing only as the mechanism by which the libration is damped. 

\section{Free Libration: The Undamped Case \label{Undamped}}

We will now consider the limit of small oscillations, which our conclusions will show to be justified.  In the undamped case, for small libration, the motion arising from the planet's torque on the moon's figure (modeled as a triaxial ellipsoid characterized by principal moments of inertia $A<B<C$) is described by \citep[e.g.,][\Eqn{}~14.3.1]{Danby88}
\begin{equation}
\ddot{\psi} = - \frac{3GM}{a^3} \left( \frac{B-A}{C} \right) \psi , 
\end{equation}
\noindent where we are assuming a circular orbit ($\ddot{\theta}=0$, and thus $\dot{\theta}=n$) of semimajor axis $a$ about a planet with mass $M$.  Forced libration due to orbital eccentricity is considered in Section~\ref{Forced}.  

This is simple harmonic motion, with a solution of the form 
\begin{equation}
\psi = \alpha \cos(\omega_0 t + \beta) , \\
\end{equation}
where $\alpha$ and $\beta$ are constants to be determined by satisfying initial conditions and
\begin{equation}
\label{SHO}
\omega_0 = \sqrt{ \frac{3GM}{a^3} \frac{B-A}{C} } \approx n \sqrt{\frac{3(B-A)}{C}} .
\end{equation}
Given our assumption of an instantaneous orbit swap (see Section~\ref{Naive}), the initial conditions are $\dot{\psi}(t=0) = n_1 - n_2$ and $\psi(t=0)=0$, resulting in
\begin{equation}
\label{UndampedPsi}
\psi = \frac{n_1 - n_2}{\omega_0} \sin \omega_0 t .
\end{equation}
\noindent Using orbital parameters from \citet{Jake08} and shape parameters from Section~\ref{Data}, listed in Table~\ref{JEInfo}, we find the amplitude of the free libration in this case is $0.022^\circ$ for Janus and $0.046^\circ$ for Epimetheus. 

\begin{table}[!b]
\begin{scriptsize}
\caption{Parameters for free libration \label{JEInfo}}
\begin{tabular} { l c c c c c c }
\hline
\hline
 & $n_1$ ($^\circ$/dy) & $n_2$ ($^\circ$/dy) & $n_1 - n_2$ ($^\circ$/dy) & $(B-A)/C$ & $\omega_0$ ($^\circ$/dy) & Amplitude ($^\circ$) \\
\hline
Janus & 518.238 & 518.346 & 0.108 & 0.100 & 284 & 0.022 \\
Epimetheus & 518.486 & 518.098 & -0.388 & 0.296 & 488 & 0.046 \\
\hline
\end{tabular}
\\Note: $n_1$ and $n_2$ are taken from \citet{Jake08}, corresponding to intervals before and after the 
\vspace{-0.08in}
\\ \indent{} 2006 orbit swap, while $(B-A)/C$ comes from Table~\ref{pct_params} below.
\vspace{-0.08in}
\\Amplitude is $\psi_0$, the value of $\psi$ when $t = 0$ in the undamped case (\Eqn{}~\ref{UndampedPsi}). 
\end{scriptsize}
\end{table}

\section{Free Libration: The Damped Case \label{Damped}}

Dissipation within the interior of the moon creates a frictional damping torque on the libration.  Parameterizing the friction as $b$, we now have the damped harmonic oscillator:
\begin{equation}
\ddot{\psi} + b\dot{\psi} + \omega_0^2 \psi = 0 .
\end{equation}
\noindent The only case in which libration proceeds is the under-damped case ($b/2\omega_0 < 1$), for which the general solution is
\begin{equation}
\psi = e^{-bt/2} \left(\alpha \cos \omega t + \beta \sin \omega t \right) , 
\end{equation}
\noindent where $\alpha$ and $\beta$ again are constants to be determined, and
\begin{equation}
\label{OmegaEqn}
\omega = \omega_0 \sqrt{1-\frac{b^2}{4\omega_0^2}} .
\end{equation}

\noindent Applying our initial conditions, as before, in the case of an instantaneous orbit swap, we obtain
\begin{equation}
\psi = \frac{n_1 - n_2}{\omega} e^{-bt/2} \sin \omega t .
\end{equation}

What is the relation between $b$ and the tidal dissipation parameter $Q$?  Following the derivation of \citet[p.161]{MD99}, the peak energy stored in the oscillator per unit mass is 
\begin{equation}
\label{E0}
E_0 = \int_0^\mathscr{A} \omega_0^2 \psi \ud \psi = \frac{1}{2} \omega_0^2 \mathscr{A}^2
\end{equation}
\noindent for instantaneous amplitude $\mathscr{A}$, and the energy dissipated per unit mass per cycle is 
\begin{equation}
\Delta E = \frac{1}{2} b (\mathscr{A} \omega)^2 \cdot \frac{2 \pi}{\omega} , 
\end{equation}
\noindent giving
\begin{equation}
Q = \frac{2 \pi E_0}{\Delta E} = \frac{\omega_0^2}{b \omega} .
\end{equation}
\noindent Substituting for $\omega$ from \Eqn{}~\ref{OmegaEqn}, we solve the resulting quadratic in $(b/\omega_0)^2$.  Rejecting the positive branch of the quadratic solution because of its unphysical behavior at large $Q$, we obtain
\begin{equation}
\label{bQ}
\left( \frac{b}{\omega_0} \right)^2 = 2 - 2 \sqrt{1-\frac{1}{Q^2}} \approx \frac{1}{Q^2} ,
\end{equation} 
\noindent where the latter approximation is valid for $Q \gg 1$.  In practice, $b \approx \omega_0 / Q$ holds for $Q \gtrsim 2$.  

We can now write the damping $e$-folding timescale, $\tau = 2/b$, as 0.40$Q$ days for Janus, and 0.24$Q$ days for Epimetheus.  
We again use $Q \sim 100$ for a solid icy body and $Q \sim 10$ for a fractured rubble-pile, but the difference in damping timescale between the two cases is now only one order of magnitude instead of two (as it was for the na\"ive treatment in Section~\ref{Naive}), as the rigidity no longer enters into our equations. 

The initial libration amplitudes in the damped case are equal to the values listed in Table~\ref{JEInfo} divided by the quantity $\sqrt{1-b^2/4\omega_0^2}$, which is very close to unity for the values of $Q$ we expect (see \Eqn{}~\ref{bQ}). 

It is worth considering whether the energy dissipated in the damping of the free libration might be a significant heat source for the moons.  From \Eqn{}~\ref{E0}, using values for $\omega_0$ and $\mathscr{A}$ given in Table~\ref{JEInfo}, and using the mean radius $\bar{R}$ (Table~\ref{JEInfo3}) to convert the angular amplitude into one in units of distance, the initial energy per unit mass stored in the oscillator is $\sim 10^{-5}$~J/kg for Epimetheus, and ten times less for Janus.  Given the idealization of our model, as stated before, this is an upper limit.  Even if this energy were dissipated very quickly, given the specific heat capacity of water ice, the moons would be heated by only $\sim 10^{-9}$~K. 

A possible refinement to this model would be a more sophisticated treatment of the mean-motion evolution during an orbit swap, perhaps in a numerical integration or by use of a polynomial fit to the mean motion from a numerical integration.  However, the current model of an instantaneous swap, when compared to any more realistic model, will maximize the resulting free-libration amplitude.  Thus our calculated amplitudes should be interpreted as upper limits, which are sufficient given our conclusion that the free librations induced on the two co-orbitals by the orbit swap are not currently observable.  

The result of this calculation is that the amplitude of the free libration excited by the orbit swap of Janus and Epimetheus is always less than $0.1^\circ$, regardless of our choice of $Q$, even in the idealized case treated here.  Furthermore, the damping timescale is only a few weeks, even in the case of very weak damping ($Q \sim 100$).  Current \Cassit{} Imaging data have neither the spatial resolution nor the temporal sampling necessary to detect the free libration. 

\section{Forced Libration \label{Forced}}

\begin{figure}[!t]
\begin{center}
\includegraphics[width=10cm]{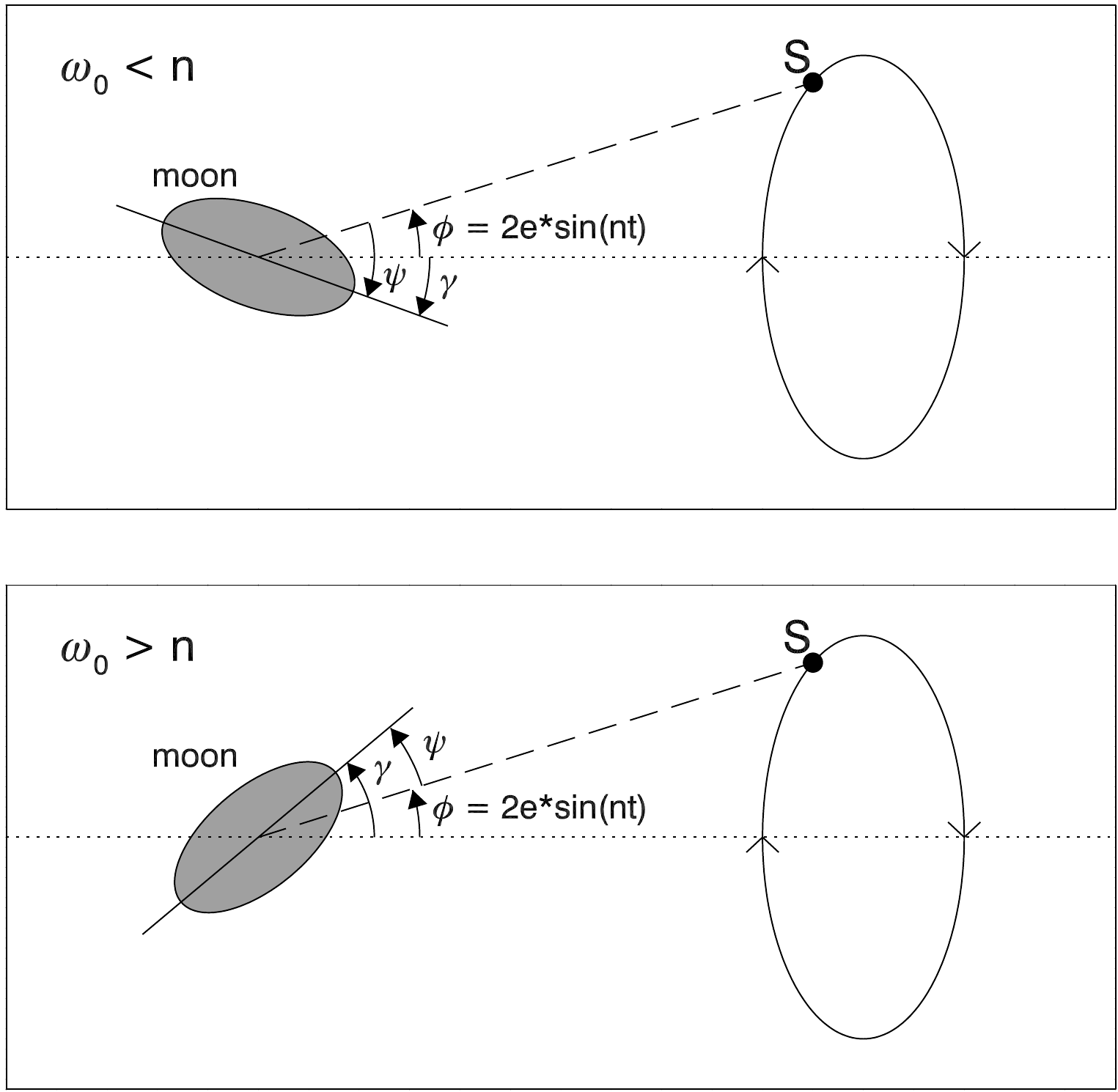}
\caption{The orientation of the moon's long axis is measured from the direction towards Saturn ($\psi$) and from the direction towards the empty focus of the moon's orbit ($\gamma$).  The latter is the deviation from synchronous rotation for $e \ll 1$.  All angles are positive in the counter-clockwise direction, so both $\psi$ and $\gamma$ are negative in the upper panel.  The optical libration angle $\phi$ is determined by the eccentricity and orbital phase.  The natural frequency $\omega_0$, which is determined by the shape parameter $(B-A)/C$ (\Eqn{}~\ref{SHO}),  determines whether the moon's long axis points away from or past the planet, as seen from the empty focus of the moon's orbit (towards which a synchronously-rotating moon would face).  The transition at $\omega_0 \sim n$ occurs when $(B-A)/C \sim 1/3$.  \label{coords_fig3a}}
\end{center}
\end{figure}

Rotational libration can also be forced by any orbital eccentricity of the moons.  We still define $\psi$ as the angle between the moon's long axis and the direction towards Saturn, and we now additionally define $\gamma$ as the angle between the moon's long axis and the empty focus of the moon's orbit (\Fig{}~\ref{coords_fig3a}).  The former is more important when considering tidal effects, while the latter is the deviation from synchronous rotation (the ``physical libration''), due to the well-known result that, to first order in eccentricity, a synchronously-rotating moon always keeps the same face towards the empty focus \citep[e.g.,][p.44]{MD99}.  The optical libration angle is $\phi = 2 e \sin nt = \gamma - \psi$.  

This problem has the solution \citep[e.g.,][Eq.~5.123]{MD99}
\begin{equation}
\label{ForcedLib}
\gamma = \frac{2 e}{1-(n/\omega_0)^2} \sin nt ,
\end{equation}
\noindent which is equivalent to
\begin{equation}
\label{ForcedLibPsi}
\psi = \frac{2 e}{(\omega_0/n)^2-1} \sin nt ,
\end{equation}
\noindent where $e$ is the eccentricity and $t$ is the time since the most recent periapse.  

If the orbit of the moon is given, the amplitude of the forced libration is then determined by the natural frequency $\omega_0$, which arises directly from the moon's shape.  Rearranging \Eqn{}~\ref{SHO}, we have
\begin{equation}
\label{Transition}
\left( \frac{\omega_0}{n} \right)^2 \approx 3 \left( \frac{B-A}{C} \right) .
\end{equation}

Most moons are close enough to spherical that $\omega_0 \ll n$, with the result that $\gamma$ approaches zero and $\psi$ approaches $-\phi$ (\Fig{}~\ref{coords_fig3a}a).  That is, the moon responds so sluggishly to Saturn's torques that it remains close to synchronous rotation, with the same face always towards the empty focus.  On the other hand, a moon might be so elongated that $\omega_0 \gg n$, with the result that $\psi$ approaches zero and $\gamma$ approaches $\phi$ (\Fig{}~\ref{coords_fig3a}b), which is to say that the moon responds so rapidly to Saturn's torques that it keeps the same face towards Saturn despite its eccentric orbit.  

The range $0 < \gamma < \phi$, which is equivalent to $-\phi < \psi < 0$, is forbidden.  That is, the moon's long axis can never point between the direction towards Saturn and the direction towards the empty focus.  We note that, due to an apparent drafting error, Figure~5.16 of \citet{MD99} illustrates this forbidden configuration.

\begin{figure}[!t]
\begin{center}
\includegraphics[width=12cm]{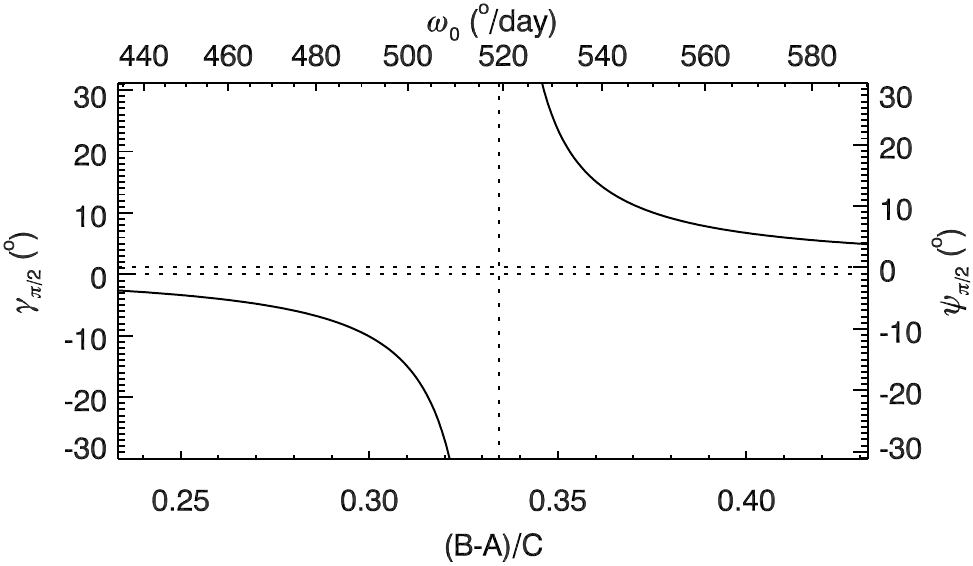}
\caption{Predicted libration amplitudes for $\gamma$ and $\psi$, calculated from \Eqn{s}~\ref{ForcedLib} and~\ref{ForcedLibPsi} with $nt = \pi/2$ for the case of Epimetheus, are plotted against the changing inertial moments (shape) of the moon.  Large amplitudes occur in the vicinity of the resonant value of the shape-derived gravity parameter, $(B-A)/C \sim 1/3$, at which the natural frequency $\omega_0$ approaches the orbital mean motion (the forcing frequency) $\bar{n} = 518.292^\circ$/day. 
\label{ForcedLibAmp}}
\vspace{-0.1in}
\end{center}
\end{figure}

\begin{table}[!b]
\begin{scriptsize}
\caption{Parameters for shape-derived prediction of forced libration \label{JEInfo2}}
\begin{tabular} { l c c c c c }
\hline
\hline
 & $\bar{n}$ ($^\circ$/dy) & $\bar{e}$ & $(B-A)/C$ & $\omega_0$ ($^\circ$/dy) & Amplitude ($^\circ$) \\
\hline
Janus & 518.292 & 0.0068 & $0.100 \pm 0.012$ & $284 \pm 17$ & $-0.33 \pm 0.06$ \\
Epimetheus & 518.292 & 0.0098 & $0.296^{+0.019}_{-0.027}$ & $488^{+15}_{-23}$ & $-8.9^{-10.4}_{+4.2}$ \\
\hline
\end{tabular}
\\Note: $\bar{n}$ and $\bar{e}$ are taken from \citet{Jake08}, while $(B-A)/C$ comes from Table~\ref{pct_params} below.
\vspace{-0.08in}
\\Amplitude is $\gamma_{\pi/2}$, the value of $\gamma$ when $nt = \pi/2$ (\Eqn{}~\ref{ForcedLib}). 
\end{scriptsize}
\end{table}

The most interesting behavior occurs near the transition between these two regimes, where $\omega_0 \sim n$, which is to say that the moon's shape is such that $(B-A)/C$ is near one-third (\Eqn{}~\ref{Transition}).  In this case, small divisors cause $\gamma$ and $\psi$ to become both very negative ($\omega_0 < n$) or both very positive ($\omega_0 > n$), as seen in \Fig{}~\ref{ForcedLibAmp}.  An extreme case of this is Hyperion, whose shape brings it close enough to $\omega_0 \sim n$ so that, in combination with its relatively high eccentricity, our assumption of low libration amplitude is violated --- as can be seen by plugging Hyperion's values, $(B-A)/C \sim 0.26$ and $e \sim 0.1$, into \Eqn{s}~\ref{ForcedLib} and~\ref{Transition}.  Instead, Hyperion has a three-dimensional rotation state characterized by chaotic tumbling \citep{Wisdom84,Wisdom87}.  As we will see, Epimetheus is close enough to this regime that its libration amplitude varies sensitively with its shape (\Fig{}~\ref{ForcedLibAmp}), allowing us to determine its gravitational moments with greater precision by measuring the libration than is achieved with direct observation of the shape.  

Preliminary calculated values of the forced libration amplitude are listed in Table~\ref{JEInfo2}, where $(B-A)/C$ is taken from the shape parameters presented in Section~\ref{Data}, and mean values of $n$ and $e$ are used since the variations are small.  For Janus, the forced libration is larger than the free libration due to the orbit swap, but is still quite small at $0.33^\circ$.  For Epimetheus, however, the amplitude of the forced libration is well into the detectable range.  The calculated value is $8.9^\circ$, though with a fairly large uncertainty since $(B-A)/C$ is not far from one-third (compare the values in Table~\ref{JEInfo2} to \Fig{}~\ref{ForcedLibAmp}).  This result will be further refined by our analysis in Section~\ref{Data}. 

\section{Data \label{Data}}
\begin{figure}[!t]
\begin{center}
\includegraphics[width=16cm]{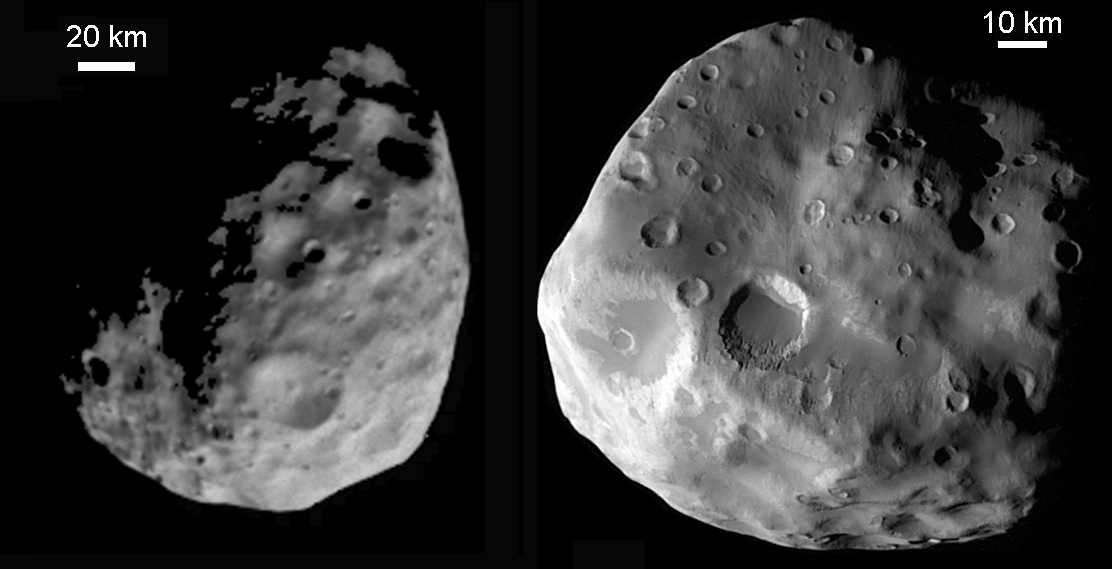}
\caption{Janus (\textit{left}) and Epimetheus (\textit{right}), from \Cassit{} images N1537923147 and N1575363491.  Note the different scale bars.  \label{JEImage}}
\vspace{-0.1in}
\end{center}
\end{figure}

Janus and Epimetheus have been imaged on a few occasions at pixel scales better than 1~km, and dozens of times at lower resolution, by the Imaging Science Subsystem \citep{PorcoSSR04} on board the \Cassit{} spacecraft (\Fig{}~\ref{JEImage}).  Details for images used in this paper are given in Tables~\ref{image_table_janus} and~\ref{image_table_epimetheus}.  From these data, a numerical model of the size and shape of each moon has been constructed using established techniques \citep{Thomas93,Simonelli93,Thomas98}.  Stereogrammetric solution of control points is the first step, and provides relative three-dimensional reference points over most of the surface (for both satellites, the solutions were closed around all longitudes).  The libration solution is part of this ``boot-strap'' process that is necessary for reducing errors in the control points and deriving an accurate shape; the libration's amplitude and phase are involved because they determine the satellite's precise orientation as seen in any image.  We obtained 66~control points on janus, and 49~on Epimetheus.  Limb and terminator positions are then used to constrain the model further, and are crucial for filling in the model where there is sparse coverage of control points.  The accuracy of the shape models is restricted by the available views and resolutions, and by the limited seasonal illumination that so far leaves small areas near the north poles poorly observed.  The estimated accuracy of a shape is calculated from the resolutions and orientations of the applied images.  Shape models for Janus and Epimetheus were given by \citet{PorcoSci07}, but are here revised with a more accurate accounting for the rotation state, as well as using additional recent data.  The improved rotational models and additional image data, for the first time, allow shape models to be useful for moment-of-inertia calculations. 

\begin{table}[!b]
\begin{scriptsize}
\caption{Observing information for images used in this paper (Janus) \label{image_table_janus}}
\begin{tabular} { c c c c c c c c c }
\hline
\hline
 & \# of & & Sub-S/C & Sub-S/C & Sub-Solar & Sub-Solar & Phase & Range \\
Image Identifier & Images & Date & Latitude & Longitude & Latitude & Longitude & Angle & (km) \\
\hline
1487417508 & & 2005-49 & 0.2 & 154.56 & -22.39 & 52.97 & 100.783 & 911572 \\
1495308742 & & 2005-140 & -27.21 & 219.96 & -21.81 & 221.74 & 5.637 & 357370 \\
1501558704 -- 58826 & 4 & 2005-213 & -20.44 & 296.1 & -20.76 & 267.0 & 27.24 & 923600 \\
1501711325 -- 11501 & 3 & 2005-214 & 11.77 & 248.4 & -20.73 & 102.1 & 146.56 & 539000 \\
1507877360 -- 78655 & 19 & 2005-286 & -0.073 & 114 & -20.13 & 8 & 105.02 & 879000 \\
1521539019 -- 39613 & 3 & 2006-79 & -0.29 & 278 & -18.16 & 230. & 50.8 & 493000 \\
1521670216 -- 70692 & 6 & 2006-80 & 0.35 & 138 & -18.15 & 296 & 152.26 & 724500 \\
1524901386 -- 01922 & 4 & 2006-118 & -0.36 & 303 & -17.85 & 240. & 63.9 & 703000 \\
1524964907 -- 66330 & 9 & 2006-119 & -0.34 & 235 & -17.84 & 264 & 33.6 & 223000 \\
1537919879 -- 23147 & 22 & 2006-268/269 & 21 & 176 & -15.78 & 227 & 61.7 & 149000 \\
1563765026 & & 2007-203 & 0.07 & 106.34 & -11.32 & 99.01 & 13.519 & 892568 \\
1582238183 -- 40507 & 10 & 2008-51 & -72.1 & 95 & -8.31 & 44 & 70.9 & 175000 \\
1589742955 -- 43095 & 2 & 2008-138 & 42.5 & 320.7 & -6.91 & 59.4 & 101.1 & 350200 \\
1590458717 -- 59027 & 4 & 2008-147 & -73.85 & 304.6 & -6.73 & 33.6 & 83.26 & 185000 \\
1590461157 & & 2008-147 & -73.05 & 317.88 & -6.73 & 47.42 & 83.43 & 213017 \\
1593508083 -- 09823 & 11 & 2008-182 & -33 -- 21 & 107 & -6.12 & 332 & 130. & 36000 \\
1594708747 -- 09210 & 2 & 2008-196 & 71.7 & 323 & -5.94 & 330 & 77.8 & 259000 \\
1602109316 & & 2008-281 & 32.79 & 85.8 & -4.86 & 84.57 & 37.668 & 1012129 \\
\hline
\end{tabular}
\\ Latitudes, longitudes, and phase angle are in degrees.  
\vspace{-0.08in}
\\ For lines referring to multiple images, variation in each parameter is in the last significant figure. 
\end{scriptsize}
\end{table}

\begin{table}[!b]
\begin{scriptsize}
\caption{Observing information for images used in this paper (Epimetheus) \label{image_table_epimetheus}}
\begin{tabular} { c c c c c c c c c }
\hline
\hline
 & \# of & & Sub-S/C & Sub-S/C & Sub-Solar & Sub-Solar & Phase & Range \\
Image Identifier & Images & Date & Latitude & Longitude & Latitude & Longitude & Angle & (km) \\
\hline
1487432013 & & 2005-49 & -0.09 & 160.66 & -22.35 & 61.37 & 98.552 & 989916 \\
1490836693 -- 36932 & 8 & 2005-89 & -0.55 & 204.4 & -21.84 & 322.1 & 115.3 & 74600 \\
1493544135 & & 2005-120 & -17.89 & 43.99 & -21.85 & 8.61 & 33.437 & 1628222 \\
1495309992 & & 2005-140 & -27.56 & 139.33 & -21.88 & 167.57 & 26.211 & 344611 \\
1500071960 -- 72512 & 6 & 2005-195 & 34.5 & 223.8 & -21.2 & 305 & 95.3 & 88000 \\
1501559484 -- 59606 & 4 & 2005-213 & -22.77 & 253.9 & -20.8 & 224.1 & 27.71 & 833300 \\
1501711887 & & 2005-214 & 13.26 & 207.29 & -20.76 & 69.79 & 138.793 & 470743 \\
1514191733 & & 2005-359 & 0.65 & 270.8 & -19.45 & 69.55 & 151.95 & 452136 \\
1516413432 & & 2006-20 & 0.01 & 236.46 & -18.84 & 81.08 & 149.366 & 1527858 \\
1521538726 -- 39613 & 3 & 2006-79 & -0.05 & 268 & -17.94 & 220. & 50.9 & 453000 \\
1524904716 -- 05315 & 5 & 2006-118 & -0.24 & 304 & -17.9 & 243 & 62.8 & 667000 \\
1560620123 & & 2007-166 & 0.78 & 140.49 & -12.03 & 180.66 & 41.888 & 1693820 \\
1575363079 -- 64219 & 14 & 2007-337 & -40.0 & 99 & -9.53 & 34 & 64.5 & 40000 \\
1596337648 -- 37691 & 2 & 2008-215 & 29.40 & 167.1 & -5.52 & 158.4 & 35.91 & 991900 \\
1599590509 & & 2008-252 & 42.79 & 240.56 & -5.38 & 223.4 & 50.625 & 859395 \\
\hline
\end{tabular}
\\ Latitudes, longitudes, and phase angle are in degrees. 
\vspace{-0.08in}
\\ For lines referring to multiple images, variation in each parameter is in the last significant figure. 
\end{scriptsize}
\end{table}

Volume and mean radius are computed directly from the numerical shape models, and gravitational moments of inertia are calculated from the models under an assumption of constant density.  An ellipsoidal fit to the numerical shape model is done only to provide the reader with a first-order estimate of the principal dimensions, and should not be used for moment calculations other than crude approximations.  

In order to compile a large suite of images, taken at different times from distinct geometries, into a single shape model, one must make an assumption about the rotation state.  Different assumed rotation states can be tested sequentially, and the best one can be determined by minimizing the residuals (measured in rms~pixels) between predicted and actual locations of the control points in the images.  While it is often sufficient to assume a constant rotation rate (which may well be synchronous rotation), perhaps with the superposition of a simple libration, the rotation states of Janus and Epimetheus cannot be described so simply.  The periodic orbit swaps change the orbital mean motion, and thus not only the synchronous rotation rate but also the frequency of the forced libration.  A free libration would further complicate the model, but we neglect free librations due to the conclusions reached in Section~\ref{Damped}. 

We addressed the problem by compiling a sequence of binary-PCK kernels in the SPICE navigation data system \citep{SPICE}.\fn{Available at \texttt{http://naif.jpl.nasa.gov/naif/toolkit.html}}  In contrast to the more common text-PCK kernels, binary-PCK kernels allow for arbitrary orientation of a given body as a function of time.  Our base state was to have the moon keep one face towards Saturn at all times ($\psi = 0$), whether before, during, or after an orbit swap.  A forced libration of any amplitude can then be superposed atop that base state.  We are thus assuming that the frequency of the forced libration responds adiabatically to the changing mean motion, an assumption we justify by noting that the orbit swap proceeds slowly (several months) compared to a single orbit (16.7~hr). 

\begin{figure}[!t]
\begin{center}
\includegraphics[width=16cm]{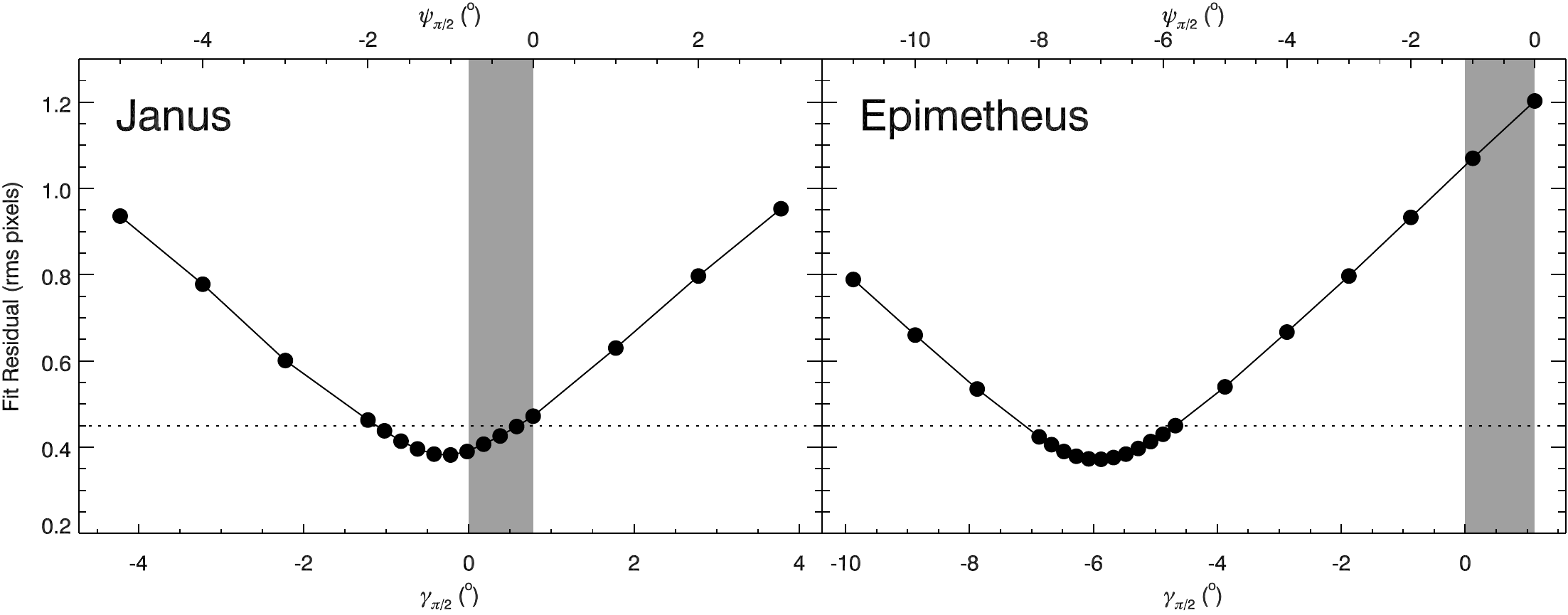}
\caption{Residuals (observed $-$ predicted image locations) for the solution of control points for Janus and Epimetheus, as a function of the amplitude of assumed forced libration (see Section~\ref{Forced}).  We used 66~control points in the fit for Janus, and 49~for Epimetheus.  The region shaded in gray, where $0 < \gamma < \phi$ and $-\phi < \psi < 0$, is dynamically forbidden.  The horizontal dotted line is the threshold used to estimate the uncertainty in the measured amplitude. 
\label{pct_data}}
\vspace{-0.1in}
\end{center}
\end{figure}

The results of this process are shown in \Fig{}~\ref{pct_data}.  For both moons, we found a clear best-fit value for the libration amplitude, with fit residuals increasing as the amplitude changes in either direction.  The best-fit residuals are 0.372 (Janus) and 0.382 (Epimetheus) rms~pixels, indicating a high-quality fit in both cases.  

We used two methods to estimate the uncertainty in the derived libration amplitude.  Firstly, we simply adopted 0.45 rms~pixels as a threshold indicating significantly reduced accuracy in the control-point solution (see the horizontal dotted line in \Fig{}~\ref{pct_data}).  The control points are nearly all crater rims; measurements are the image locations of the center of an ellipse fit to the rim.  On objects with well-known spins and no morphological complications, solutions typically can reach $\sim 1/3$~pixels.  The overall solution here is not likely to reach this common level due to inclusion of some low-resolution views where large less-circular craters are of necessity included.  Residuals above $\sim 0.4$~pixels indicate some problem with input geometry or the necessity to include irregularly-shaped craters.  Here 0.45~pixels is a safe measure of poor input geometry, that is, incorrect assumed libration amplitudes.  Making use of a quadratic fit to the points within the threshold, we find $\gamma_{\pi/2} = -0.3^\circ \pm 0.9^\circ$ (Janus) and $\gamma_{\pi/2} = -5.9^\circ \pm 1.2^\circ$ (Epimetheus).  As a second method of estimating the uncertainty, we take the best solution, and allow half the control points to have the best-fit error, and the other half an error of 0.5 pixels, and use the resulting average rms residuals as the error estimate.  This method gives results nearly identical to the first method. 

A crude check can be made by considering the surface resolution of good images that affect the solution.  For Epimetheus, this is 0.44~km/pixel, so that 0.45~pixels corresponds to 0.2$^\circ$ on the surface of a body of mean radius 58.1~km.  Likewise, for Janus, the result is 0.3$^\circ$.  Thus the estimated uncertainties from \Fig{}~\ref{pct_data} are reasonable and conservative. 

\begin{table}[!b]
\begin{scriptsize}
\caption{Parameters for best-fit numerical shape models \label{pct_params}}
\begin{tabular} { l c c c c c c c c }
\hline
\hline
 & $A/MR^2$ & $B/MR^2$ & $C/MR^2$ & $(B-A)/C$ & $a$ (km) & $b$ (km) & $c$ (km) & $\bar{R}$ (km) \\
\hline
Janus & 0.360 & 0.407 & 0.470 & $0.100 \pm 0.012$ & 101.5 $\pm$ 1.9 & 92.5 $\pm$ 1.2 & 76.3 $\pm$ 1.2 & 89.5 $\pm$ 1.4 \\
Epimetheus & 0.328 & 0.469 & 0.476 & $0.296^{+0.019}_{-0.027}$ & 64.9 $\pm$ 2.0 & 57.0 $\pm$ 3.7 & 53.1 $\pm$ 0.7 & 58.1 $\pm$ 1.8 \\
\hline
\end{tabular}
\\ Principal moments of inertia ($A$, $B$, and $C$) are calculated under an assumption of constant density.  
\vspace{-0.08in}
\\ Mean radius $\bar{R}$, the radius of a sphere of equivalent volume, is derived directly from the numerical shape model. 
\vspace{-0.08in}
\\ Axis lengths $a$, $b$, and $c$ come from an ellipsoidal fit. 
\end{scriptsize}
\end{table}

The parameters of the best-fit shape model are given in Table~\ref{pct_params}.  The measured values given here for mean radius $\bar{R}$ and moment of inertia ratio $(B-A)/C$ have been retroactively used for the preliminary calculations in the preceding sections (Tables~\ref{JEInfo3} through~\ref{JEInfo2}).  

The moment of inertia ratio $(B-A)/C$ is taken directly from the numerical shape model.  Though we give axis lengths ($a$, $b$, $c$) of an ellipsoid fit to the numerical shape model, the moment ratio estimated from the ellipsoid, $(B-A)/C \sim (a^2-b^2)/(a^2+b^2)$, gives a highly divergent value in the case of Epimetheus, whose shape in fact deviates significantly from an ellipsoid.  We estimate uncertainties in the model moments by adding and subtracting bulges to the shape model, consistent with constraints from image data, for six comparison models.  The maximum and minimum $(B-A)/C$ are then used as estimated bounds on moment ratios.  

We are now in a position to compare the measured rotation state to that expected from calculations based on the shape model.  With our measured values of the amplitude $\gamma_{\pi/2}$ (that is, the value of $\gamma$ when $nt = \pi/2$; see \Eqn{}~\ref{ForcedLib}) and the known orbital eccentricity and mean motion, we re-derive the moment of inertia ratio. 

For Janus, the measured libration amplitude implies $(B-A)/C = 0.09^{+0.11}_{-0.09}$, which is consistent with, but clearly less precise than, the shape-derived value.  However, it is interesting that, despite the large uncertainty, the measured $\gamma_{\pi/2} = -0.3^\circ$ is very close to the shape-derived prediction of $-0.33^\circ$ (Table~\ref{JEInfo2}).  This would tend to validate the assumption of constant density, which was employed in order to derive gravitational moments from the shape model.  On the other hand, we note that Janus' sub-Saturn point (when $nt = 0$) is offset from the axis of the minimum moment of inertia ($A$), by $5.2^\circ \pm 1^\circ$ in longitude and $2.3^\circ \pm 1^\circ$ in latitude.  This may indicate that Janus does in fact have some internal density asymmetries, though the latitude offset at least may be due simply to vertical librations (equivalent to a non-zero obliquity and precession), which have not been taken into account in the current model.  Unlike Epimetheus, Janus has a significant value of $(C-B)/A$, making it more responsive to vertical perturbations. 

For Epimetheus, the measured libration amplitude implies $(B-A)/C = 0.280^{+0.008}_{-0.011}$, significantly improving upon the precision of the shape-derived value.  The libration-measured value is in agreement with the shape-derived estimate, which has a fairly large uncertainty, so we cannot constrain the possibility of internal density asymmetries without further improvements to the shape model.  Furthermore, in contrast to Janus, Epimetheus' sub-Saturn point for $nt=0$ deviates from the minimum moment of inertia axis by less than $1^\circ$ in both latitude and longitude.  

\section{Summary and Conclusions \label{Conclusions}}

\begin{table}[!b]
\begin{scriptsize}
\caption{Summary of results \label{SumTable}}
\begin{tabular} { l l c c }
\hline
\hline
 & & Janus & Epimetheus \\
\hline
\multicolumn{2}{l}{Free libration, shape-derived prediction} & & \\
 & Amplitude & 0.022$^\circ$ & 0.046$^\circ$ \\
 & Damping time, solid ($Q \sim 100$) & 40 days & 24 days \\
 & Damping time, fractured ($Q \sim 10$) & 4 days & 2.4 days \\
\multicolumn{2}{l}{Optical libration, $\phi_{\pi/2} = 2 e$} & 0.8$^\circ$ & 1.1$^\circ$ \\
\multicolumn{2}{l}{Forced libration, $\gamma_{\pi/2}$} & & \\
 & Amplitude, shape-derived prediction & $-0.33^\circ \pm 0.06^\circ$ & $-8.9^\circ$ $^{-10.4^\circ}_{+4.2^\circ}$ \\
 & Amplitude, measured & $-0.3^\circ \pm 0.9^\circ$ & $-5.9^\circ \pm 1.2^\circ$ \\
\multicolumn{2}{l}{Moment of inertia ratio $(B-A)/C$} & & \\
 & Shape-derived (assuming constant density) & $0.100 \pm 0.012$ & $0.296^{+0.019}_{-0.027}$ \\
 & Inferred from measured libration & $0.09^{+0.11}_{-0.09}$ & $0.280^{+0.008}_{-0.011}$ \\
\hline
\end{tabular}
\end{scriptsize}
\end{table}

Although the free libration, the ``ringing'' of the moons' rotation states due to their periodic orbit swap, is a tantalizing target, our calculations show that it is not detectable by \Cassit{} Imaging.  Not only is the amplitude too small ($<0.1^\circ$), even with some idealized assumptions, but the decay rate is too fast (a few weeks, at most). On the other hand, our calculations indicate that the forced libration, which arises on any non-spherical moon with an eccentric orbit, should be barely detectable for Janus and quite large for Epimetheus.  

Our analysis of \Cassit{} Imaging data, combined with a model of the forced libration, yields improved shape models and (for Epimetheus) the third clear detection of forced libration in the solar system (after Earth's Moon and Mars' Phobos).  Furthermore, Epimetheus' measured amplitude of forced libration allows significant improvements in the precision of the shape-derived estimate of the ratio $(B-A)/C$ of principal moments of inertia.  

Although the measured forced-libration amplitude for Janus is indistinguishable from zero (synchronous rotation) due to its uncertainty, it is consistent with the shape-derived estimate of $(B-A)/C$.  Moreover, we have detected a permanent offset of several degrees between the shape model for Janus and the observed equilibrium rotation state.  

We have particular interest in comparing the orientations and/or magnitudes of the principal moments of inertia inferred from analysis of their rotational librations with those inferred from analysis of their shapes.  Since the latter involves an assumption of homogeneous density, we account for any discrepancies (such as the offset in orientation detected in Janus) by suggesting that internal density asymmetries exist within the moon.  These may take the form of internal voids \citep{GT69} and/or of a highly porous accreted mantle atop a monolithic core \citep{PorcoSci07}. 

The principal step that might be taken to improve this method would be to consider vertical librations (i.e., non-zero obliquity and precession).  Neglecting vertical perturbations is certainly justified in the case of Epimetheus, which is quite prolate in shape ($A < B \sim C$), though including them may improve the results for Janus.  

According to a companion paper by \citet{Morrison09}, sets of parallel grooves observed on the surfaces of Epimetheus and Phobos, but not on moons for which significant libration can be ruled out, also occur on Pandora and cannot be ruled on on Prometheus.  We plan to apply our method to these and other moons in the Saturn system, in the hopes of detecting forced librations that can be used to probe the interiors of orbiting objects. 

\acknowledgements{We thank B.~Carcich for his assistance in implementing the creation of binary-PCK kernels in SPICE.  We thank P.~Nicholson, M.~Hedman, C.~Murray, F.~Nimmo, and an anonymous reviewer for helpful comments, and the \Cassit{} Imaging Team for their assistance in planning and processing the observations.  We acknowledge funding from the Cassini Project, from NASA's Cassini Data Analysis Program (NNX08AQ72G), and from NASA's Planetary Geology \& Geophysics program (NNX08AL25G).}

\bibliographystyle{apalike}
\bibliography{bibliography}

\end{document}